\documentclass[11pt,twoside]{article}
\usepackage{asp2014}

\aspSuppressVolSlug
\resetcounters

\begin{document}

\title{Agile methodologies in teams with highly creative and autonomous members}

\author{Sergi~Blanco-Cuaresma,
        Alberto~Accomazzi,
        Michael~J.~Kurtz,
        Edwin~Henneken,
        Carolyn~S.~Grant,
        Donna~M.~Thompson,
        Roman~Chyla,
        Stephen~McDonald,
        Golnaz~Shapurian,
        Timothy~W.~Hostetler,
        Matthew~R.~Templeton,
        Kelly~E.~Lockhart,
        and Kris~Bukovi
  }
\affil{Harvard-Smithsonian Center for Astrophysics, 60 Garden Street, Cambridge, MA 02138, USA \email{sblancocuaresma@cfa.harvard.edu}}
% remove/add as you need

% remove/add authors as you need
\paperauthor{Sergi~Blanco-Cuaresma}{sblancocuaresma@cfa.harvard.edu}{0000-0002-1584-0171}{Harvard-Smithsonian Center for Astrophysics}{HEAD}{Cambridge}{MA}{02138}{USA}
\paperauthor{Alberto~Accomazzi}{aaccomazzi@cfa.harvard.edu}{0000-0002-4110-3511}{Harvard-Smithsonian Center for Astrophysics}{HEAD}{Cambridge}{MA}{02138}{USA}
\paperauthor{Michael~J.~Kurtz}{kurtz@cfa.harvard.edu}{0000-0002-6949-0090}{Harvard-Smithsonian Center for Astrophysics}{HEAD}{Cambridge}{MA}{02138}{USA}
\paperauthor{Edwin~A.~Henneken}{ehenneken@cfa.harvard.edu}{0000-0003-4264-2450}{Harvard-Smithsonian Center for Astrophysics}{HEAD}{Cambridge}{MA}{02138}{USA}
\paperauthor{Carolyn~S.~Grant}{cgrant@cfa.harvard.edu}{0000-0003-4424-7366}{Harvard-Smithsonian Center for Astrophysics}{HEAD}{Cambridge}{MA}{02138}{USA}
\paperauthor{Donna~M.~Thompson}{dthompson@cfa.harvard.edu}{0000-0001-6870-2365}{Harvard-Smithsonian Center for Astrophysics}{HEAD}{Cambridge}{MA}{02138}{USA}
\paperauthor{Roman~Chyla}{rchyla@cfa.harvard.edu}{0000-0003-3041-2092}{Harvard-Smithsonian Center for Astrophysics}{HEAD}{Cambridge}{MA}{02138}{USA}
\paperauthor{Stephen~McDonald}{stephen.mcdonald@cfa.harvard.edu}{0000-0003-1270-0605}{Harvard-Smithsonian Center for Astrophysics}{HEAD}{Cambridge}{MA}{02138}{USA}
\paperauthor{Golnaz~Shapurian}{gshapurian@cfa.harvard.edu}{0000-0001-9759-9811}{Harvard-Smithsonian Center for Astrophysics}{HEAD}{Cambridge}{MA}{02138}{USA}
\paperauthor{Timothy~W.~Hostetler}{thostetler@cfa.harvard.edu}{0000-0001-9238-3667}{Harvard-Smithsonian Center for Astrophysics}{HEAD}{Cambridge}{MA}{02138}{USA}
\paperauthor{Matthew~R.~Templeton}{matthew.templeton@cfa.harvard.edu}{0000-0003-1918-0622}{Harvard-Smithsonian Center for Astrophysics}{HEAD}{Cambridge}{MA}{02138}{USA}
\paperauthor{Kelly~E.~Lockhart}{kelly.lockhart@cfa.harvard.edu}{0000-0002-8130-1440}{Harvard-Smithsonian Center for Astrophysics}{HEAD}{Cambridge}{MA}{02138}{USA}
\paperauthor{Kris~Bukovi}{kbukovi@cfa.harvard.edu}{0000-0002-5827-2434}{Harvard-Smithsonian Center for Astrophysics}{HEAD}{Cambridge}{MA}{02138}{USA}

% leave these next few aindex lines commented for the editors to enable them. Use Aindex.py to generate them for yourself.
% first presenting author should be the first entry for bold-facing the author index page-reference
%\aindex{Blanco-Cuaresma,~S.}
%\aindex{Accomazzi,~A.}
%\aindex{Kurtz,~M.~J.}
%\aindex{Henneken,~E.}
%\aindex{Grant,~C.~S.}
%\aindex{Thompson,~D.~M.}
%\aindex{Chyla,~R.}
%\aindex{McDonald,~S.}
%\aindex{Shapurian,~G.}
%\aindex{Hostetler,~T.~W.}
%\aindex{Templeton,~M.~R.}
%\aindex{Lockhart,~K.~E.}
%\aindex{Bukovi,~K.}
%\aindex{Rapport,~N.}
% remove/add as you need

% leave the ssindex lines commented for the editors to enable them, use Index.py to suggest yours
%\ssindex{FOOBAR!conference!ADASS 2019}
%\ssindex{FOOBAR!organisations!ASP}

% leave the ooindex lines commented for the editors to enable them, use ascl.py to suggest yours
%\ooindex{FOOBAR, ascl:1101.010}
  
\begin{abstract}
The Agile manifesto encourages us to value individuals and interactions over processes and tools, while Scrum, the most adopted Agile development methodology, is essentially based on roles, events, artifacts, and the rules that bind them together (i.e., processes). Moreover, it is generally proclaimed that whenever a Scrum project does not succeed, the reason is because Scrum was not implemented correctly and not because Scrum may have its own flaws. This grants irrefutability to the methodology, discouraging deviations to fit the actual needs and peculiarities of the developers. In particular, the members of the NASA ADS team are highly creative and autonomous whose motivation can be affected if their freedom is too strongly constrained. We present our experience following Agile principles, reusing certain Scrum elements and seeking the satisfaction of the team members, while rapidly reacting/keeping the project in line with our stakeholders expectations.
\end{abstract}

\section{Introduction}

The NASA Astrophysics Data System \citep[ADS;][]{2000A&AS..143...41K} is a bibliographic database for astronomical research. ADS contains now more than 14 million records and 100 million citations, its content has significantly increased since its conception \citep{2000A&AS..143..111G} and it includes software and data citations \citep{2015scop.confE...3A}. The original architecture \citep{2000A&AS..143...85A} and user interface \citep{2000A&AS..143...61E} have been recently replaced, adding more advanced functionality \citep{2015ASPC..495..401C, 2015ASPC..492..189A, 2018AAS...23136217A} and taking advantage of a new cloud infrastructure \citep{2019ASPC..523..353B}.

The project is managed following an agile methodology strongly based on Scrum but with adaptations to the peculiarities of ADS team, which is described in the next section. 

\articlefigure[width=.5\textwidth]{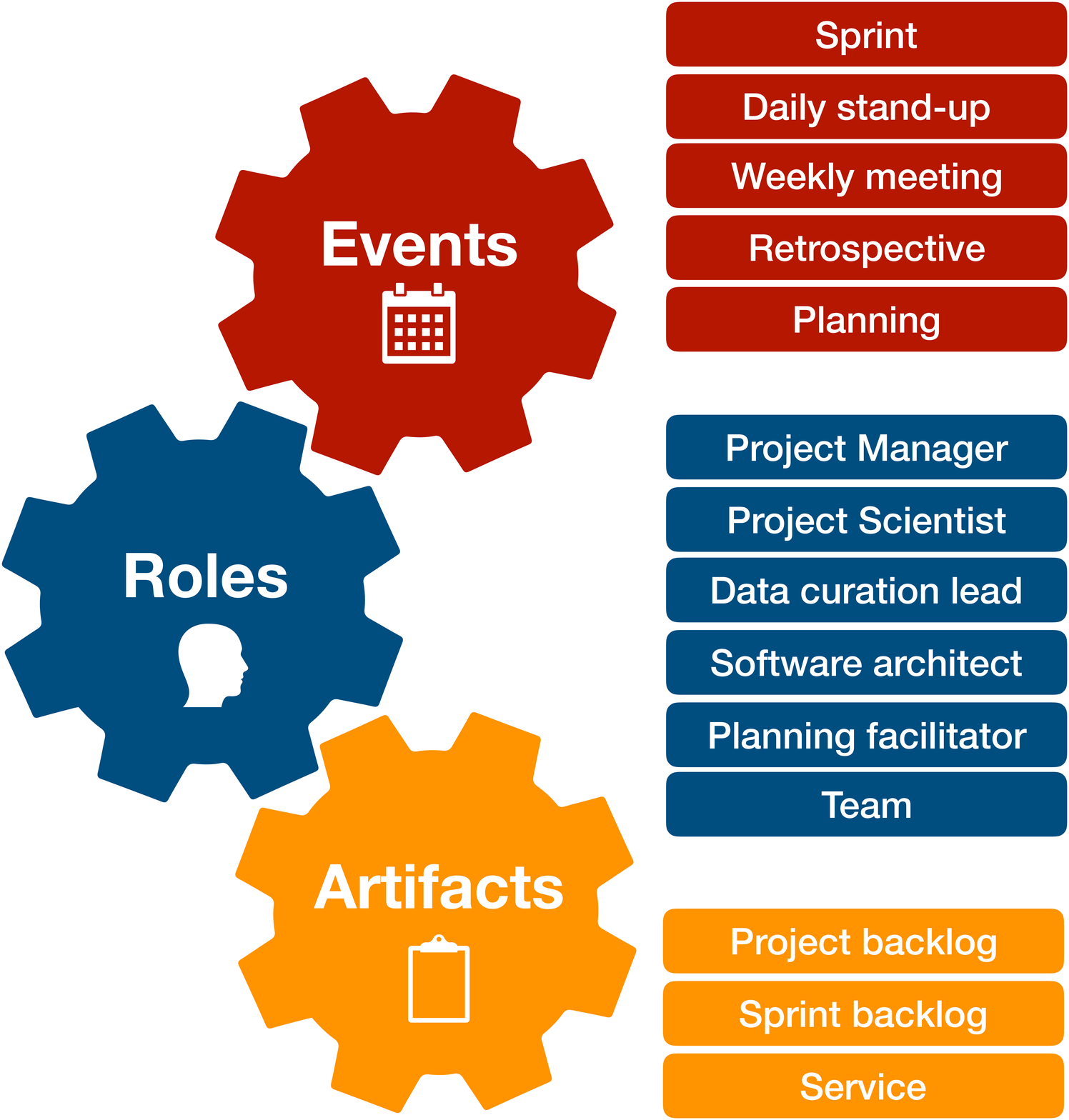}{fig1}{Main components of the agile methodology followed by NASA ADS.}

\section{Agile methodology}

ADS is fortunate to have team members with heterogeneous backgrounds who can contribute to the project with different perspectives: data curators, software engineers / scientists with experience in the public and private sector, and astronomers. Most of the members are highly motivated to propose and lead the development of new features, perform studies and carry out research activities. This leads to a culture of proactive involvement in the development process, which would not fit well with a more industrialized approach of creating tasks/users stories and forcing members to pick the first one on the top of a prioritized list. A potential drawback of not following certain strict rules that methodologies such as Scrum proposes is that the development process might be slightly slower (e.g., it might happen that eventually there are team members not working on the top 1 priority), but the benefit of adapting the methodology to the ADS particularities is an increase in professional satisfaction and a lower degree of member rotation (which, in the long run, almost always have a bigger impact on the efficiency than the previously mentioned drawback).

The main key elements of the methodology are shown in Fig.~\ref{fig1} and summarized in the next subsection.

\subsection{Events}

After multiple iterations, we have converged to sprints with a duration of 3 weeks. During that period, we do daily stand-up meetings of maximum 30 minutes where we share what we did the previous day, what we plan today and blockers that we may be facing. In addition, we have weekly meetings with the full team (developers and curators) to present progress and coordinate efforts. At the end of the sprint, we do a one hour retrospective to identify what can be done differently, followed by one hour planning meeting where the developers have a considerable amount of freedom to choose what is going to be their goal during the following sprint.

\subsection{Roles}

The project manager sets the global goals and priorities for the project in coordination with the project scientist, the curation lead, the software architect and with the advice of multiple team members who are developers/curators and astronomers, thus they can provide feedback from a user-centric point of view. The planning facilitator structures the meetings where some of these conversations happen, helps to find consensus and maintains documents with the agreements.

\subsection{Artifacts}

The planning facilitator keeps a backlog of major technical tasks that need to be done to accomplish the goals agreed by the project manager, software architect, developers and curators. During the sprint planning, the team members can select the tasks they would like to work on, propose new tasks or redefine existing ones to include more details and address shortcomings. The result of this group effort is the ADS system, which is constantly monitored and updated with periodic releases.

\section{Summary}

Within the agile methodologies, Scrum seems to be the leading one, but at the same time many authors are raising concerns such as:

\begin{itemize}
    \item Scrum is fragile since large numbers of teams with intelligent software developers do not seem to succeed to implement it successfully\footnote{Source: \url{http://www.dennisweyland.net/blog/?p=43}\\Discussion: \url{https://news.ycombinator.com/item?id=20017854} }
    \item The Agile Industrial Complex (i.e., agile consulting firms) has the habit of imposing processes that go against the agile principles\footnote{Source: \url{https://martinfowler.com/articles/agile-aus-2018.html}\\Discussion: \url{https://news.ycombinator.com/item?id=18272594} }
    \item Agile teams do not usually inspect and adapt but limit themselves to adding increments from backlogs as if they were working in a factory or assembly line\footnote{Source: \url{https://medium.com/columbus-egg/dear-agile-im-tired-of-pretending-d39ab6a12003}\\Discussion: \url{https://news.ycombinator.com/item?id=20325096} }
\end{itemize}

The ADS team tries to overcome these concerns (among others) by following an agile methodology strongly based on scrum but adapted to the peculiarities of the team members. This search for a better way to organize the work and boost individual motivation does not end in the components presented here but it will keep evolving with time, adapting to new circumstances in a continuous improving process.

\bibliography{P9-20}

% if we have space left, we might add a conference photograph here. Leave commented for now.
% \bookpartphoto[width=1.0\textwidth]{foobar.eps}{FooBar Photo (Photo: Any Photographer)}

\end{document}